\documentclass[12pt]{article}
\usepackage{a4,amssymb,verbatim,amsmath}
\numberwithin{equation}{section}

\newtheorem{theorem}{Theorem}[section]
\newtheorem{lemma}[theorem]{Lemma}
\newtheorem{remark}[theorem]{Remark}

\newtheorem{proposition}[theorem]{Proposition}
\newtheorem{definition}[theorem]{Definition}

\newtheorem{ex}{Example}[section]

\newtheorem{ass}{Assumption}[section]

\numberwithin{equation}{section}

\begin{document}

\def\dx#1{{\partial \over \partial#1}}
\def\dh#1{\mathop {#1}\limits_{h}}
\def\dhp#1{\mathop {#1}\limits_{+h}}
\def\dhm#1{ \mathop{#1}\limits_{-h}}
\def\dphh#1{ \mathop{#1}\limits_{h \bar h}}
\def\da#1{ \mathop{#1}\limits_{+\tau}}
\def\db#1{ \mathop{#1}\limits_{-\tau}}
\def\dc#1{ \mathop{#1}\limits_{\pm \tau}}
\def\dd#1{ \mathop{#1}\limits_{+h}}
\def\df#1{ \mathop{#1}\limits_{-h}}
\def\dpm#1{ \mathop{#1}\limits_{\pm h}}
\def\dg#1{ \mathop{#1}\limits_{\pm h}}
\def\dh#1{ \mathop{#1}\limits_ h}

\def\sso#1{\ensuremath{\mathfrak{#1}}}

\newcommand{\ddt}{\partial \over \partial t}
\newcommand{\ddx}{\partial \over \partial x}
\newcommand{\ddy}{\partial \over \partial y}
\newcommand{\ddyy}{\partial \over \partial y'}

\begin{center}
{\Large {\bf Invariance and first integrals of \\
continuous and discrete  Hamiltonian equations}}
\end{center}

\begin{center}

{\large  Vladimir Dorodnitsyn}$^{*}$ {\large and Roman
Kozlov}$^{\ddag}$

\medskip
\hspace{1.5 cm}

${}^{*}$ Keldysh Institute of Applied Mathematics,
Russian Academy of Science,\\
Miusskaya Pl.~4, Moscow, 125047, Russia; \\
E-mail address: dorod@spp.Keldysh.ru \\

${}^{\ddag}$ Department of Finance and Management Science, Norwegian School of Economics \\
and Business Administration, Helleveien 30, N-5045, Bergen, Norway; \\
E-mail address: Roman.Kozlov@nhh.no \\

\bigskip
\bigskip

  { \bf 15.05.2009}
  \end{center}

\bigskip

\bigskip
\begin{center}
{\bf Abstract}
\end{center}
\begin {quotation}
In this paper we consider the relation between symmetries and first
integrals for both continuous canonical Hamiltonian equations and
discrete Hamiltonian equations.
 We observe that canonical Hamiltonian equations can be obtained
by variational principle from an action functional and consider
invariance properties of this functional as it is done in Lagrangian
formalism. We rewrite the well--known Noether's identity in terms of
the Hamiltonian function and symmetry operators. This approach,
based on symmetries of the Hamiltonian action, provides a simple and
clear way to construct first integrals of Hamiltonian equations
without integration. A discrete analog of this identity is
developed. It leads to a relation between symmetries and first
integrals for discrete Hamiltonian equations that can be used to
conserve structural properties of Hamiltonian equations in numerical
implementation.  The results are illustrated by a number of examples
for both continuous and discrete Hamiltonian equations.
\end{quotation}

\section{Introduction}
\label{Intro}

It has been known since E.~Noether's fundamental work that
conservation laws of differential equations are connected with
their symmetry properties~\cite{Noe}. For convenience we present
here some well--known results (see also, for example,
\cite{Abraham, Goldstein, Arnold})
for both Lagrangian and Hamiltonian
approaches to conservation laws (first integrals).

Let us consider the functional
\begin{equation}   \label{lagyr}
\mathbb{L}(u) =  \int_{\Omega} L ( x,u,u_1) dx ,
\end{equation}
where
$ x=(x^1,x^2,...,x^m)$ are independent variables,
$ u=(u^1,u^2,...,u^n)$ are dependent variables,
$  u_1=(u^k_i)$ are all first order derivatives
$ u^k_i= {\partial u^k \over \partial x^i  }$,
$\Omega$ is a domain in $ \mathbb{R}^m$
and $L ( x,u,u_1)$ is a {\it first order} Lagrangian.
The functional~(\ref{lagyr}) achieves its extremal values when $u(x)$
satisfies the  Euler--Lagrange equations
\begin{equation}   \label{extyr}
{ \delta L \over \delta u^k } = {\partial L \over \partial u^k  }
- D_i \left( {\partial L \over \partial u^k_i } \right) = 0,
\qquad
k=1,...,n,
\end{equation}
where
$$
D_i = {\ddx^i} + u^k_i { \partial \over\partial u^k } +
u^k_{ji} { \partial \over\partial u^k_j } + \cdots ,
\qquad i = 1,...,m
$$
are total differentiation operators with respect to independent variables
$x^i$.
Here and below we assume summation over repeated indexes.
Note that equations~(\ref{extyr}) are second order PDEs.

We consider a Lie point transformation group $G$ generated by
the infinitesimal operator
\begin{equation}  \label{symmetry0}
X = \xi^i ( x,u)  { \partial \over \partial x^i } + \eta ^k  (x,u)
{ \partial \over \partial u^k } +  ...,
\end{equation}
where dots mean an appropriate prolongation of the operator on partial
derivatives~\cite{Ovs, Olver, Ibr, Blu}.
The group $G$ is called a variational symmetry of the functional
$\mathbb{L}(u) $ if and  only if the Lagrangian satisfies~\cite{Noe}
\begin{equation} \label{cond}
  X(L) + L D_i( \xi^i) = 0,
\end{equation}
where  $ X$ is the first prolongation, i.e. the prolongation  of the
vector field $X$ on the first derivatives $u^k_i$. We will actually
need a weaker invariance condition than given by Eq.~(\ref{cond}).
The vector field $X$ is a divergence symmetry of the functional
$\mathbb{L}(u) $ if there exist functions $V^i(x,u,u_1)$, $i = 1, ..., m$
such that~\cite{Bess} (see also~\cite{Olver, Ibr, Blu})
\begin{equation} \label{cong}
 X(L) + L D_i( \xi^i) = D_i(V^i) .
\end{equation}

An important result for us is the following:
If $X$ is a variational symmetry
 of the functional $\mathbb{L}(u) $, it is also a
symmetry of the corresponding Euler--Lagrange equation. The
symmetry group of Eqs.~(\ref{extyr}) can of course be larger than
the group generated  by  variational and divergence symmetries
of the Lagrangian.

 Noether's theorem~\cite{Noe}  states that for a Lagrangian satisfying
the condition~(\ref{cond}) there exists a conservation law of the
Euler--Lagrange equations~(\ref{extyr}):
\begin{equation} \label{conserv}
D_i \left(\xi^i L + (\eta^k - \xi^j u^k_j){ \partial L\over \partial
u^k_i } \right) = 0.
\end{equation}
This result can be generalized:
If $X$ is a divergence symmetry of the functional $\mathbb{L}(u) $,
i.e. equation~(\ref{cong}) is satisfied, then there exists a
conservation law
\begin{equation} \label{cdrtre}
D_i \left(\xi^i L + (\eta^k - \xi^j u^k_j){ \partial L\over \partial
u^k_i }-V^i \right) = 0
\end{equation}
of the corresponding Euler--Lagrange equations.

The strong version of the Noether's theorem~\cite{Ibr} states that
there exists a conservation law of the Euler--Lagrange
Eqs.~(\ref{extyr}) in the form~(\ref{conserv}) if and only if
the condition~(\ref{cond}) is satisfied on the solutions of
Eqs.~(\ref{extyr}).

\bigskip

In the present paper we  are interested in canonical Hamiltonian
equations
\begin{equation}  \label{canonical}
\dot{q} ^i = { \partial H \over  \partial {p}_i } , \qquad
\dot{p}_i = - { \partial H \over  \partial {q} ^i }, \qquad
i = 1, ... , n .
\end{equation}
These equations can be obtained
by the variational principle
from the action functional
\begin{equation}  \label{principe}
\delta   \int_{t_1} ^{t_2}
\left(
 p_i   \dot{q} ^i  - H ( t, {\bf q} , {\bf p} )
\right) dt = 0
\end{equation}
in the phase space $( {\bf q}, {\bf p} )$,
where ${\bf q} = ( q ^1, q ^2,..., q ^n ) $,
${\bf p} = ( p_1, p_2, ..., p_n ) $ (see, for
example,~\cite{Gelfand, Marsden}).

Let us note that the canonical Hamiltonian equations~(\ref{canonical})
can be obtained by action of the variational operators
\begin{equation}  \label{varoperator1}
{ \delta \over \delta p _i }  = { \partial \over \partial p_i } - D
{ \partial \over \partial \dot{p}_i } , \qquad  i = 1, ..., n  ,
\end{equation}
\begin{equation}  \label{varoperator2}
{ \delta \over \delta
q ^i } = { \partial \over \partial q  ^i } - D { \partial \over
\partial \dot{q} ^i } , \qquad i = 1, ..., n  ,
\end{equation}
where $D$ is the operator of total differentiation with respect to
time
\begin{equation}  \label{derivative}
D = { \partial  \over  \partial t }
+     \dot{q} ^i { \partial \over \partial q ^i }
+    \dot{p}_i { \partial \over  \partial p_i } +
...,
\end{equation}
on the function
$$
  p_i   \dot{q} ^i  - H ( t,  {\bf q} , {\bf p} ) .
$$
The Legendre transformation relates Hamiltonian and Lagrange
functions
\begin{equation}  \label{Legendre}
 L  ( t, {\bf q} , { \dot{\bf q}} )
 = p_i   \dot{q} ^i  - H ( t, {\bf q} , {\bf p} ),
\end{equation}
where ${\bf p}={ \partial L\over \partial \dot{\bf q} }, {\dot{\bf
q}}={\partial H\over \partial {\bf p} }$. It makes it possible to
establish the equivalence of the Euler--Lagrange and Hamiltonian
equations~\cite{Arnold}.
Indeed, from Euler--Lagrange equations
for one independent variable ($m=1$)
\begin{equation}   \label{euler}
{ \delta L \over \delta q^i } = {\partial L \over \partial q^i  }
- D \left( {\partial L \over \partial \dot{q}^i } \right) = 0,
\qquad
i = 1, ..., n
\end{equation}
we can obtain the canonical Hamiltonian equations~(\ref{canonical})
using the Legendre transformation. It should be noticed that the
Legendre transformation is not a point transformation. Hence, there is no
conservation of Lie group properties of the corresponding
Euler--Lagrange equations and Hamiltonian equations within the class
of point transformations.

\bigskip

Lie point symmetries in the space $(t,{\bf q},{\bf p})$ are
generated by operators of the form
\begin{equation}  \label{symmetry}
X = \xi ( t, {\bf q}, {\bf p} )  { \partial \over \partial t }
+ \eta ^i  ( t, {\bf q},  {\bf p} )  { \partial \over \partial q ^i }
+  \zeta _i  ( t, {\bf q},  {\bf p})  { \partial \over \partial p_i  }.
\end{equation}
Standard approach to symmetry properties of the Hamiltonian
equations
is to
consider so called {\it Hamiltonian symmetries}~\cite{Olver}. In
the case of canonical Hamiltonian equations these are
the evolutionary  ($\xi = 0$) symmetries~(\ref{symmetry})
\begin{equation}  \label{evolsymmetry}
\bar{X}  = \eta ^i  ( t, {\bf q},  {\bf p} )  { \partial \over \partial q ^i }
+   \zeta _i  ( t, {\bf q},  {\bf p})  { \partial \over \partial p_i  }
\end{equation}
with
\begin{equation}  \label{integration}
\eta ^i   = {  \partial I  \over \partial p_i  } ,
\qquad
\zeta ^i  = -  {  \partial I  \over \partial q ^i  },
\qquad
i = 1, ..., n
\end{equation}
for some function $ I (t,{\bf q},{\bf p})$,
namely,  symmetries of the form
\begin{equation}    \label{Hsymmetry}
\bar{X}_I  =   {  \partial I  \over \partial p_i  }
 { \partial \over \partial q ^i  }
 -  {  \partial I  \over \partial q ^i  }
 { \partial \over \partial p_i  }  .
\end{equation}
These symmetries  are restricted to the phase space
$( {\bf q},{\bf p} )$  and are generated by the function
$I=I(t, {\bf q}, {\bf p})$.
For symmetry~(\ref{Hsymmetry}) the independent variable $t$ is
invariant and plays a role of a parameter.

Noether's theorem (Theorem 6.33 in~\cite{Olver})
relates Hamiltonian symmetries of the
Hamiltonian equations with their first integrals.
Restricted to the case of the canonical Hamiltonian equations
it can be formulated as follows:

\begin{proposition}   \label{Prop}
An evolutionary vector field $\bar{X}$ of the form~(\ref{evolsymmetry})
generates a Hamiltonian symmetry group of the canonical
Hamiltonian system~(\ref{canonical}) if and only if there exists a
first integral  $ I (t,{\bf q},{\bf p})$ so that $ \bar{X} = \bar{X} _I $ is
the corresponding Hamiltonian vector field.
Another function $ \tilde{I} (t,{\bf q},{\bf p})$
determines the same Hamiltonian symmetry if and only if $
\tilde{I} = I + F(t) $ for some time-dependent function $ F(t) $.
\end{proposition}



Thus, we obtain that the Hamiltonian symmetry determines a first
integral of the canonical Hamiltonian equations up to some
time-dependent function, which can be found with the help of these
equations.
This approach has two disadvantages. First, some transformations loose
their geometrical sense if considered in evolutionary form~(\ref{Hsymmetry}).
Second, there is a necessity of integration to find
first integrals with the help of~(\ref{integration}).
In this approach it is also not clear why
some point symmetries of Hamiltonian equations yield integrals,
while others do not.

In the present paper we will consider symmetries of the general form~(\ref{symmetry}),
which are not restricted to the phase space and
can also transform $t$. In contrast to the Hamiltonian symmetries in
the form~(\ref{Hsymmetry}) the underlying symmetries have a clear
geometric sense in finite space and do not require integration
to find first integrals. We will provide a Hamiltonian
version of the Noether's theorem (in the strong formulation)  based
on a newly established Hamiltonian identity, which is an analog of
well--known Noether's identity for the Lagrangian approach. The
Hamiltonian identity links directly an invariant Hamiltonian
function with first integrals of the canonical Hamiltonian
equations. This approach provides a simple and clear way to
construct first integrals by means of merely algebraic manipulations
with symmetries of the action functional. The approach will be
illustrated on a number of examples, including equations of the
three-dimensional Kepler motion.

\bigskip

The paper is organized as follows:
In section~\ref{invariance_section}
we introduce a definition of an invariant Hamiltonian and establish the
necessary and sufficient condition for $H$ to be invariant.
Section~\ref{identity} contains the main propositions of present
paper: Lemma~\ref{lemmaidentity} introduces a new identity,
which is used in Theorem~\ref{firstintegral}
to formulate the necessary and sufficient condition for existence of
first integrals of Hamiltonian equations (Hamiltonian version of
Noether's theorem in the strong formulation). In
section~\ref{invariance}, Lemma~\ref{lemmavariation}
introduces two more identities,
which are used in Theorem~\ref{invarianceofvariations}
to formulate necessary and sufficient
conditions for the canonical Hamiltonian equations to be invariant.
Section~\ref{applications} contains example ODEs
which are considered as both Euler--Lagrange equations and
canonical Hamiltonian equations. In particular, we
consider the equations of Kepler motion. In
section~\ref{discreteHamiltonian} we present discrete Hamiltonian
equations. Their symmetries and first integrals are shown to be
related in the same way as those for the continuous canonical
Hamiltonian equations. Final section~\ref{conclusion} contains
concluding remarks.

\section{Invariance of elementary Hamiltonian action }
\label{invariance_section}

As an analog of the Lagrangian elementary action~\cite{Olver, Ibr}
we consider the Hamiltonian elementary  action
\begin{equation}  \label{action}
  p_i  d{q} ^i   -  H dt ,
\end{equation}
which can be invariant or not with respect to a group
generated by an operator of the form~(\ref{symmetry}).

\begin{definition}
We call a Hamiltonian function invariant with respect to
a symmetry operator~(\ref{symmetry}) if the elementary action~(\ref{action}) is an
invariant of the group generated by this operator.
\end{definition}

\begin{theorem} \label{Haminvariance}
A Hamiltonian is invariant with respect to a group generated by
the operator~(\ref{symmetry}) if and only if the following condition holds
\begin{equation}  \label{invar}
    {  \zeta} _i   \dot{q} ^i  +  p _i  D( {  \eta}^i  )
-  X ( H ) - H  D( \xi )  = 0.
\end{equation}
\end{theorem}

\noindent {\it Proof.}$\ $
The invariance condition follows directly from  the action of
the operator $X$ prolonged on the differentials  $dt$
and  $d{ q ^i }$, $i = 1, ..., n$~\cite{Ibr}:
\begin{equation}  \label{prol}
X = \xi ( t, {\bf q}, {\bf p} )  { \partial \over \partial t }
+ \eta ^i  ( t, {\bf q},  {\bf p} )  { \partial \over \partial q ^i }
+ \zeta _i  ( t, {\bf q},  {\bf p})  { \partial \over \partial p_i }
+ D(\xi ) dt  { \partial \over \partial (dt) }
+ D(\eta ^i )  dt  { \partial \over \partial {(dq ^i)} } .
\end{equation}
Application of~(\ref{prol})
to the Hamiltonian elementary action  (\ref{action}) yields
$$
X \left(      p_i  d q ^i  - H dt \right) =
\left(   {  \zeta} _i \dot{q} ^i  + p_ i  D( {  \eta}^i  )
-  X ( H ) - H D(\xi ) \right)  dt =  0.
$$
\hfill $\Box$

\begin{remark}
From the relation
\begin{equation}  \label{Legen}
 L  ( t, {\bf q} , { \dot{\bf q}} )dt
 = p_i   d{q} ^i  - H ( t, {\bf q} , {\bf p}  ) dt
\end{equation}
it follows  that if a Lagrangian is invariant with respect to a group of
Lie point transformations, then the Hamiltonian
is also invariant with respect to the same group (of point transformations).
The converse statement is false.
For example, symmetries providing components of Runge--Lenz
vector as first integrals of Kepler motion are point symmetries
in Hamiltonian framework (point~\ref{secKepler}).
However, they are generalized symmetries in
Lagrangian framework~\cite{Olver}.
\end{remark}

\noindent The proof follows from the action of operator~(\ref{symmetry})
on relation~(\ref{Legen}).

\begin{remark}
The operator of total differentiation~(\ref{derivative})
applied to Hamiltonian $H$ and considered on the solutions of
Hamiltonian equations~(\ref{canonical})  coincides with partial
differentiation with respect to time:
\begin{equation}  \label{diff}
\left.  D(H) \right|
_{ \dot{\bf q} = H_{\bf p}, \ \dot{\bf p} = - H_{\bf q} }
= \left[ { \partial H \over  \partial t }
+     \dot{q} ^i  { \partial H \over \partial q ^i  }
+     \dot{p}_ i { \partial H \over \partial p_i  }
  \right]
  _{ \dot{\bf q} = H_{\bf p}, \ \dot{\bf p} = - H_{\bf q} }
= { \partial H \over  \partial t }.
\end{equation}
\end{remark}

\section{The Hamiltonian identity and Noether--type theorem}
\label{identity}

Now we can relate conservation properties of the canonical
 Hamiltonian equations to the invariance of the Hamiltonian function.

\begin{lemma}   \label{lemmaidentity}
The identity
\begin{equation}  \label{ident}
\begin{array}{c}
{\displaystyle
   \zeta _i  \dot{q} ^i  + p _ i   D( \eta  ^i )
-  X ( H ) - H  D( \xi )
\equiv   \xi \left( D(H) - { \partial H \over  \partial t } \right)  } \\
\\
{\displaystyle
-   \eta^i
    \left(  \dot{p}_i  +  { \partial H \over  \partial q  _i } \right)
+  \zeta _i
    \left(  \dot{q} ^i  - { \partial H \over  \partial p_ i } \right)
+ D \left[    p _i  \eta ^i  - \xi H  \right]  } \\
\end{array}
\end{equation}
is true for any smooth function $H=H(t,{\bf q},{\bf p} )$.
\end{lemma}

\noindent {\it Proof.}$\ $
The identity can be established by direct calculation.
\hfill $\Box$

\medskip

We call this identity the {\it Hamiltonian identity}. This identity
makes it possible to develop the following
result.

\begin{theorem}   \label{firstintegral}
The canonical Hamiltonian equations~(\ref{canonical})
possess a first integral of the form
\begin{equation}  \label{integral}
I =     p_i  \eta ^i  - \xi H
\end{equation}
if and only if the Hamiltonian function is invariant with
respect to the operator~(\ref{symmetry}) on the solutions of
the equations~(\ref{canonical}).
\end{theorem}

\noindent {\it Proof.}$\ $
The result follows from the identity~(\ref{ident}).
\hfill $\Box$

\medskip

Theorem~\ref{firstintegral} corresponds to the strong version
of the Noether theorem (i.e. necessary and sufficient  condition) for
invariant Lagrangians and Euler--Lagrange  equations~\cite{Ibr}.

\begin{remark}    \label{corollfirst}
Theorem~\ref{firstintegral} can be generalized on
the case of the divergence invariance of the Hamiltonian action
\begin{equation}  \label{divinvaraince2}
    \zeta _i \dot{q} ^i  + p_ i  D( \eta ^i )
-  X ( H ) - H  D ( \xi )  = D( V ),
\end{equation}
where $V  = V ( t,{\bf q} ,{\bf p}) $.
If this condition holds on the solutions
of the canonical Hamiltonian equations~(\ref{canonical}), then
there is a first integral
\begin{equation}  \label{integ1}
I =      p_i  \eta ^i  - \xi H - V .
\end{equation}
\end{remark}

\section{Invariance of canonical Hamiltonian equations}
\label{invariance}

In the Lagrangian framework, the variational principle
provides us with the Euler--Lagrange equations. It is known that the
invariance of the  Euler--Lagrange equations follows from the
invariance of the action integral. The following Lemma~\ref{lemmavariation} and
Theorem~\ref{invarianceofequations} establish the sufficient conditions for canonical
Hamiltonian equations to be invariant.

\begin{lemma}   \label{lemmavariation}
The following identities are true for any smooth function
$H=H(t,{\bf q},{\bf p} )$:

$$
{ \delta \over \delta p _j }
\left( {  \zeta} _i   \dot{q} ^i  + p _i  D( {  \eta}^i  )
- X ( H)  - H  D( \xi ) \right)
$$
\begin{equation}  \label{conse1}
\equiv
D ( \eta^j  ) -  \dot{q} ^j  D ( \xi )
- X \left( { \partial H \over \partial p_ j  } \right)
+  { \partial \xi \over \partial p_j }
\left(  D(H) - { \partial H \over  \partial t } \right)
\end{equation}
$$
-    { \partial  \eta ^i  \over \partial p_j }
   \left(  \dot{p}_i  +  { \partial H \over  \partial q ^i  } \right)
+  \left(  { \partial  \zeta _i  \over   \partial  p_j }
    + \delta_{ij} D( \xi ) \right)
   \left( \dot{q}^i  - { \partial H \over  \partial p_i  } \right)  ,
\qquad j = 1, ..., n ,
$$

\medskip

$$
{\delta \over \delta q ^j } \left( {  \zeta} _i   \dot{q} ^i
+  p _i  D( {  \eta}^i  ) - X ( H)  - H  D( \xi ) \right)
$$
\begin{equation}  \label{conse2}
\equiv
 - D ( \zeta _j  ) +  \dot{p} _j  D ( \xi )
 -  X \left( { \partial H \over \partial q_j  } \right)
 +  { \partial \xi  \over \partial q ^j }
   \left( D(H) - { \partial H \over  \partial t }  \right)
\end{equation}
$$
-   \left(  { \partial  \eta ^i \over \partial q ^j }
    +  \delta_{ij} D(\xi) \right)
  \left(  \dot{p}_i  +  { \partial H \over  \partial q ^i  } \right)
+   { \partial  \zeta _i  \over \partial  q ^j }
  \left(  \dot{q} ^i  - { \partial H \over  \partial p_i  }  \right) ,
 \qquad
j = 1, ..., n ,
$$
where the notation $  \delta_{ij}  $ stands for the Kronecker symbol.
\end{lemma}

\noindent {\it Proof.}$\ $
The identities can be easily obtained by direct computation.
\hfill $\Box$

\begin{theorem}   \label{invarianceofequations}
If a Hamiltonian is invariant with
respect to the symmetry~(\ref{symmetry}),
then the canonical Hamiltonian equations~(\ref{canonical})
are also invariant.
\end{theorem}

\noindent {\it Proof.}$\ $
For invariance of the canonical Hamiltonian equations (\ref{canonical})
we need the equations
$$
D ( \eta ^j  ) -  \dot{q} ^j  D ( \xi ) =
X \left( { \partial H \over \partial p_ j  } \right) ,
\qquad
j = 1, ..., n
$$
$$
D ( \zeta _j  ) -   \dot{p} _j  D ( \xi ) =
 -  X \left( { \partial H \over \partial q ^j  } \right),
\qquad
j = 1, ..., n
$$
to hold on the solutions of the Hamiltonian equations~\cite{Olver}.
These conditions follow from
the identities ~(\ref{conse1}) and~(\ref{conse2}).
\hfill $\Box$

\begin{remark}
The statement of Theorem~\ref{invarianceofequations} remains valid if
we consider divergence symmetries of the Hamiltonian,
i.e. condition~(\ref{divinvaraince2}), because the term $D(V)$
belongs to the kernel of the variational operators~(\ref{varoperator1}),(\ref{varoperator2}).
\end{remark}

The invariance of the Hamiltonian is a {\it sufficient condition}
for the canonical Hamiltonian equations to be invariant.
The symmetry group of  the canonical Hamiltonian equations can of course
be larger than that of the Hamiltonian.
The following Theorem~\ref{invarianceofvariations}
establishes the {\it necessary and sufficient} conditions
for canonical Hamiltonian equations to be invariant.

\begin{theorem}   \label{invarianceofvariations}
Canonical Hamiltonian equations~(\ref{canonical})
are  invariant with respect to the symmetry~(\ref{symmetry})
{if and only if} the following conditions are
true (on the solutions of  the canonical Hamiltonian equations):
\begin{equation}  \label{new1}
\left.
{ \delta \over \delta p _j } \left( {  \zeta} _i   \dot{q} ^i
+ p _i  D( {  \eta}^i  ) - X ( H)  - H  D( \xi )
\right)
\right|
 _{ \dot{\bf q} = H_{\bf p}, \ \dot{\bf p} = - H_{\bf q} }
=0 ,
\qquad  j = 1, ..., n ,
\end{equation}
\begin{equation}  \label{new2}
\left.
{\delta \over \delta q ^j } \left( {  \zeta} _i   \dot{q} ^i
+  p _i  D( {  \eta}^i  ) - X ( H)  - H  D( \xi )
\right)
\right|
 _{ \dot{\bf q} = H_{\bf p}, \ \dot{\bf p} = - H_{\bf q} }
=0  ,
\qquad  j = 1, ..., n .
\end{equation}
\end{theorem}

\noindent {\it Proof.}$\ $
The statement follows from the identities~(\ref{conse1}) and~(\ref{conse2}).
\hfill $\Box$

\medskip

It should be noted that conditions~(\ref{new1}) and~(\ref{new2}) are true for all
symmetries of canonical Hamiltonian equations. But not all of those
symmetries yield the "variational integral" of these conditions, i.e.
$$
\left.
\left( {  \zeta} _i   \dot{q} ^i
+ p _i  D( {  \eta}^i  ) - X ( H)  - H  D( \xi )
\right)
\right|
 _{ \dot{\bf q} = H_{\bf p}, \ \dot{\bf p} = - H_{\bf q} }
=0,
$$
which gives first integrals  in accordance with
Theorem~\ref{firstintegral}. {\it That is why not all symmetries of
the canonical Hamiltonian equations provide first integrals}. In the
next section we illustrate the theorems, given above, on a number of
examples.

\section{Applications}
\label{applications}

In this section we provide examples how to find first integrals
with the help of symmetries.

\subsection{A scalar ODE.}

 As the first example we consider the second-order ODE
\begin{equation} \label{equ1}
 \ddot{u} = \frac{1}{u^{3}},
\end{equation}
which admits Lie algebra $L_3$ with basis operators
\begin{equation} \label{ope3}
 X_1 = \dx t , \qquad X_2 = 2t \dx t + u\dx u ,  \qquad
X_3 =t^2 \dx t + tu \dx u.
\end{equation}

\subsubsection{Lagrangian approach}

The Lagrangian function
\begin{equation} \label{L1}
L ( t, u, \dot{u})  =  {1 \over 2} \left(  \dot{u}^2 - \frac{1}{u^2} \right) ,
\end{equation}
which provides equation (\ref{equ1}) as its Euler--Lagrange equation,
is invariant with respect to  $X_1$ and $ X_2$. Therefore, by means of
 Noether's theorem there exist first integrals
\begin{equation} \label{int2}
J_{1} =   - {1 \over 2} \left(  \dot{u}^2 + \frac{1}{u^2} \right) ,
\qquad
J_{2} =  u  \dot{u} -  t \left(  \dot{u}^2 + \frac{1}{u^2} \right) .
\end{equation}
The action of the third operator $X_3$ yields the divergence
invariance condition
\begin{equation} \label{div2}
X {L} +  {L} D (\xi)  =  u  \dot{u} = D \left( { u^2 \over 2}  \right).
\end{equation}
Due to the divergence invariance of the Lagrangian
we can find the following first integral
\begin{equation} \label{int3}
J_{3} =  - { 1 \over 2}
\left( {t^2 \over u^2} + ( u - t \dot{u} ) ^2 \right) .
\end{equation}
Alternatively, one can find the last integral from another
Lagrangian function
\begin{equation} \label{L2}
\tilde{L} ( t, u, \dot{u} )  = \left( \frac{u}{t}-  \dot{u}  \right) ^2 - \frac{1}{u^2},
\end{equation}
which is exactly invariant with respect to  $X_3$.

It should be mentioned that independence of first integrals
obtained with the help of the Noether theorem
is guarantied only in the case when there is
one Lagrangian which is invariant with respect to all symmetries. This
condition is broken in the considered example.
Therefore, the integrals obtained are not independent.
Integrals~(\ref{int2}),(\ref{int3}) are connected
by the relations
\begin{equation} \label{connect}
4 J_{1} J_{3} -  J_{2} ^2= 1.
\end{equation}
Thus, any two integrals among~(\ref{int2}),(\ref{int3}) are
independent. Putting  $J_1 = A/2$, $J_2 = B $ and excluding $ \dot{u} $,
we find the general solution of the equations (\ref{equ1}) as
\begin{equation} \label{sol3}
A u^{2} +  ( A t - B )^{2} + 1 = 0 .
\end{equation}

\subsubsection{Hamiltonian framework}

Let us transfer the preceding example into the Hamiltonian framework.
We change variables
$$
q = u,
\qquad
p = { \partial {L} \over  \partial  \dot{u}  }=  \dot{u}  .
$$
The corresponding Hamiltonian is
\begin{equation} \label{fun}
H (t,q,p) =  \dot{u} { \partial {L} \over  \partial \dot{u} } - {L}
=  { 1 \over 2} \left(  p^2  + { 1 \over  q^2 } \right)  .
\end{equation}

The Hamiltonian equations
\begin{equation}  \label{ca}
\dot{q}  =  p  ,
\qquad
\dot{p} = \frac{1}{q^3}
\end{equation}
admit symmetries
\begin{equation} \label{opera}
 X_1 = \dx t ,
 \qquad
 X_2 = 2t \dx t + q \dx q  -p \dx p ,
 \qquad
 X_3 =t^2 \dx t + tq \dx q  + ( q - tp) \dx p .
\end{equation}

We check invariance of $H$ in accordance with Theorem~\ref{Haminvariance}
and find that condition~(\ref{invar}) is satisfied  for the operators $X_1$ and $X_2$.
Using Theorem~\ref{firstintegral},  we calculate the corresponding
first integrals
\begin{equation} \label{integ2}
I_{1} = - H  =  -  { 1 \over 2} \left(  p^2  + { 1 \over  q^2 } \right) ,
\qquad
I_{2} = pq -  t \left(  {p^2}   + { 1 \over  q^2 } \right).
\end{equation}
For the third symmetry operator the Hamiltonian is divergence
invariant with $V_3 = q^2 / 2  $.
In accordance with Remark~\ref{corollfirst},
it yields the following conserved quantity
\begin{equation} \label{integral3}
I_{3} =  - { 1 \over 2}
\left( {t^2 \over q^2} + ( q - t p  ) ^2 \right) .
\end{equation}
Note that no integration is needed to provide solutions of~(\ref{ca}).
As we noted before in the Lagrangian case
only two first integrals are functionally independent.
Putting  $I_1 = A/2$ and $I_2 = B $,
we  find the solution (\ref{ca}) as
\begin{equation} \label{sol3h}
A q^{2}  +  ( A t -  B  )^{2} + 1  = 0 ,
\qquad
p = { B -  At \over q}.
\end{equation}

\subsection{Repulsive one-dimensional motion.}

As another example of an ODE we consider one-dimensional motion in
the Coulomb field (the case of a repulsive force):
\begin{equation} \label{equ2}
 \ddot{u}  = \frac{1}{u^{2}},
\end{equation}
which admits Lie algebra $L_2$ with basis operators
\begin{equation} \label{opera3}
 X_1 = \dx t ,
\qquad
X_2 = 3t \dx t + 2u\dx u .
\end{equation}

\subsubsection{Lagrangian approach}

The Lagrangian function
\begin{equation} \label{Lag}
L ( t, u, \dot{u} )  = { \dot{u} ^2 \over 2 }  - \frac{1}{u}
\end{equation}
is invariant only with respect to  $X_1$.
Therefore,  Noether's theorem yields the only first integral
\begin{equation} \label{inte}
J_{1} =  { \dot{u} ^2 \over 2 }  + { 1 \over  u }.
\end{equation}
In this case the Euler--Lagrange equation admits two symmetries
while the Lagrangian is invariant  with respect to one symmetry
operator only.

\subsubsection{Hamiltonian framework}

We change variables
$$
q=u,
\qquad
p={ \partial {L} \over  \partial \dot{u} }= \dot{u}
$$
and find the Hamiltonian function
\begin{equation} \label{hamilt}
H (t, q, p) =  \dot{u} { \partial {L} \over  \partial \dot{u} } - {L}
=  \frac{p^2}{2} + { 1 \over  q } .
\end{equation}

The Hamiltonian equations have the form
\begin{equation}  \label{candy}
\dot{q} =  p ,
\qquad
\dot{p} = \frac{1}{q^2} .
\end{equation}
We rewrite symmetries in the canonical variables as the following
algebra $L_2$:
\begin{equation} \label{operato}
 X_1 = \dx t ,
\qquad
X_2 = 3t \dx t + 2q \dx q  - p  \dx p .
\end{equation}
The invariance of Hamiltonian condition~(\ref{invar}) is satisfied
for operator $X_1$ only. Applying Theorem~\ref{firstintegral},
we calculate the corresponding first integral
\begin{equation} \label{integ3}
I_{1} = - H = - \left(  \frac{p^2}{2} + { 1 \over  q } \right) .
\end{equation}
Application of operator $X_2$ to the Hamiltonian action gives
\begin{equation} \label{qu}
 {  \zeta}   \dot{q}   + p  D( {  \eta} ) - X  ( H) - H  D( \xi )
 = p \dot{q} -\left( \frac{p^2}{2} + { 1 \over  q }\right)
 \neq 0.
\end{equation}
Meanwhile,  in accordance with Theorem~\ref{invarianceofvariations} we have
$$
\left.
{ \delta \over \delta p } \left( {  \zeta}   \dot{q}  + p
D( {  \eta}  ) - X ( H)  - H  D( \xi )
\right)
\right|
_{ \dot{q} =  p , \ \dot{p} = \frac{1}{q^2}  }
= ( \dot{q} - p ) |
_{ \dot{q} =  p , \ \dot{p} = \frac{1}{q^2}  }
= 0 ,
$$
$$
\left.
{\delta \over \delta q } \left( {  \zeta}  \dot{q}  +  p
  D( {  \eta}  ) - X ( H)  - H  D( \xi )
\right)
\right|
_{ \dot{q} =  p , \ \dot{p} = \frac{1}{q^2}  }
= \left. \left(  - \dot{p} + { 1 \over q^2 } \right)
\right|
_{ \dot{q} =  p , \ \dot{p} = \frac{1}{q^2}  }
=0.
$$
We will show below that there exists a second integral of non-local
character.

\bigskip

It was shown in~\cite{Dor} that Eq.~(\ref{equ2}) can be linearized by
a contact transformation. For  equations~(\ref{candy}) this
transformation is the following
\begin{equation} \label{change}
p(t) = P(s) ,  \qquad Q^2(s)= \frac{2}{q(t)} ,
 \qquad  dt = -
\frac{4}{Q^3} ds.
\end{equation}
The new Hamiltonian
\begin{equation} \label{hamilton}
H (s, Q, P ) =   {1 \over 2} ( P^2  +  Q^2 )
\end{equation}
corresponds to the linear equations
\begin{equation}  \label{canHam}
  { dQ \over ds } =  P   , \qquad  { dP \over ds } = - Q,
\end{equation}
which describe the one-dimensional harmonic oscillator.
These  equations have  two independent first  integrals
\begin{equation} \label{integrals}
I_{1} =  {1 \over 2} ( P^2 +  Q^2 ) ,
\qquad
I_{2} = \arctan
\left( \frac{P}{Q}\right) + s ,
\end{equation}
which let us write down the general solution of the equations (\ref{canHam}) as
\begin{equation} \label{sine}
Q =  A \sin s + B \cos s, \qquad P= A \cos s - B \sin s,
\end{equation}
where $ A$ and $B$ are arbitrary constants.
Applying the transformation~(\ref{change}) to integral $I_{2}$
we find the  non-local integral for Eqs.~(\ref{candy})
\begin{equation} \label{integra}
I_2 ^* = \arctan \left( \frac{p\sqrt q}{\sqrt 2}\right) -
   \frac{1}{\sqrt 2}  \int_{t_0}^t \frac{dt}{ q^{3/2} }.
\end{equation}

\subsection{Kepler motion}
\label{secKepler}

Kepler's problem is a special case of the two-body problem, in which the two bodies interact by a central force
that varies in strength as the inverse square
of the distance between them~\cite{Abraham, Goldstein, Arnold}.
The three-dimensional Kepler motion of a body in Newton's
gravitational field is given by the equations
\begin{equation}  \label{Kepler}
  \dot{\bf q}  =  {\bf p}  ,
\qquad
  \dot{\bf p}  = - { K ^2 \over r  ^3  }  {\bf q} ,
\qquad
r = | {\bf q} | ,   \qquad   {\bf q} , {\bf p}  \in  \mathbb{R} ^3 ,
\end{equation}
where $K$ is a constant,  with the initial data
$$
{\bf q} ( 0 ) = {\bf q} _0 , \qquad
{\bf p} ( 0 ) = {\bf p} _0  .
$$
These equations are Hamiltonian.
They are defined by the Hamiltonian function
\begin{equation}  \label{hamil}
H ( { \bf  q }, {\bf p } )
= { 1  \over 2 }  | { \bf  p } | ^2  -  { K ^2 \over r   }  .
\end{equation}

Among symmetries admitted by the equations~(\ref{Kepler})
there are
$$
\begin{array}{l}
{\displaystyle
X_0 = { \partial \over \partial t} ,
\qquad
X_1 = 3t  { \partial \over \partial t}
+ 2 q ^i { \partial \over \partial q ^i}
- p_i { \partial \over \partial p_i}  ,} \\
\\
{\displaystyle
X_{ij} = - q ^j { \partial \over \partial q ^i}
+  q ^i { \partial \over \partial q ^j}
- p_j { \partial \over \partial p_i}
+  p_i { \partial \over \partial p_j} ,
\qquad   i \neq j    ,} \\
\\
{\displaystyle
Y_l =  ( 2 q ^l p_k - q^k p_l - ( {\bf q},  {\bf p} ) \delta _{lk} )
 { \partial \over \partial q ^k}  } \\
\\
{\displaystyle
+ \left(
   p_l p_k -  ( {\bf p},  {\bf p} ) \delta _{lk}
- { K^2 \over r^3 }  (  q^l q^k -  ( {\bf q},  {\bf q} ) \delta _{lk}  )
  \right)
{ \partial \over \partial p_k} ,
\qquad
 l = 1,2,3 , } \\
\end{array}
$$
where $( {\bf f},  {\bf g} ) = {\bf f}^T {\bf g} $
is scalar product of vectors.

The Hamiltonian function~(\ref{hamil}) is invariant for symmetries
$X_0$ and $X_{ij} $. Theorem~\ref{firstintegral} makes it possible
 to find the first integral for symmetry $X_0$
$$
I_1 = - H ,
$$
which represents the conservation of the energy in Kepler motion.
For symmetries  $X_{ij}$ we obtain the first integrals
$$
I_{ij} = q ^i p_j  -  q ^j p_i , \qquad i \neq j ,
$$
which are components of the angular momentum
\begin{equation}  \label{moment}
{\bf L} ( { \bf  q }, {\bf p } )  = {\bf q} \times {\bf p}  .
\end{equation}
Conservation of the angular momentum shows that
the orbit of motion of a body lies in a fixed plane
perpendicular to the constant vector $ {\bf L} $.
It also follows that in this plane the position
vector $  {\bf q} $ sweeps out equal areas in equal time intervals,
so that the sectorial velocity is constant~\cite{Arnold}.
Therefore, Kepler's second law can be considered as a geometric
reformulation of the conservation of angular momentum.

The scaling symmetry $ X_1$ is not a Noether symmetry
(neither variational, nor divergence symmetry)
and does not lead to a conserved quantity.

For each of symmetries    $Y_{l}$ the Hamiltonian is
divergence invariant with functions
$$
V_l = q ^l  \left(  ( {\bf p},  {\bf p} ) + { K^2 \over r } \right)
- p_l  ( {\bf q},  {\bf p} )  , \qquad  l  = 1, 2, 3.
$$
Hence, the operators   $Y_{l}$  yield the first integrals
$$
I_{l} = q ^l  \left(  ( {\bf p},  {\bf p} ) - { K^2 \over r } \right)
- p_l  ( {\bf q},  {\bf p} ) ,
\qquad  l = 1, 2, 3,
$$
which are components of the Runge--Lenz vector
\begin{equation}
{ \bf A}  ( { \bf  q }, {\bf p } ) =
{\bf p} \times {\bf L} - { K^2   \over r }  {\bf q} =
{ \bf q }  \left(  H ( { \bf  q }, {\bf p } )
+  { 1 \over 2 }  | {\bf p } | ^2   \right)
- { \bf p } ( {\bf q } ,  {\bf p } )  .
\end{equation}
Physically, vector ${ \bf A}$ points along the major axis of
the conic section determined by the orbit of the body. Its
magnitude determines the eccentricity~\cite{Thirring}.

Let us note that not all first integrals are independent. There
are two relations between them given by the equations
$$
 { \bf A} ^2 - 2 H  {\bf L}^2  =  K ^4  \qquad
\mbox{and}
\qquad
(  { \bf A}  ,  { \bf L } )  = 0 .
$$

The two-dimensional Kepler motion can be considered in a similar way.
Let us remark that symmetries and first integrals
of the two-dimensional Kepler
motion can be obtained by restricting the symmetries and first integrals of the
three-dimensional Kepler motion to the space $(t, q ^1, q ^2, p_1, p_2)$.
As the conserved quantities of the two-dimensional Kepler
motion one obtains the energy
$$
H ( { \bf  q }, {\bf p } )
= { 1  \over 2 }  | { \bf  p } | ^2  -  { K ^2 \over r   },
\qquad r = | { \bf  q } | ,
\qquad { \bf  q } = ( q ^1, q ^2 ) ,
\quad {\bf p } = ( p_1, p_2 ) ,
$$
one component of the angular momentum
$$
L_3  = { q ^1 } { p_2 }  -  { q ^2 } { p_1 }
$$
and
two components of the  Runge--Lenz vector
$$
A_1 = q ^1  \left(  H ( { \bf  q }, {\bf p } )
+  { 1 \over 2 }  | {\bf p } | ^2   \right)
-  p_1  ( {\bf q } ,  {\bf p } )  .
$$
$$
A_2 = q ^2  \left(  H ( { \bf  q }, {\bf p } )
+  { 1 \over 2 }  | {\bf p } | ^2   \right)
-  p _2  ( {\bf q } ,  {\bf p } )  .
$$
There is one relation between these conserved quantities, namely
$$
A_1 ^2 +  A_2  ^2 - 2 H  L_3 ^2  =  K ^4   .
$$

Further restriction to the one-dimensional Kepler motion leaves only one
first integral, which is the Hamiltonian function.

\section{First integrals of discrete Hamiltonian equations}
\label{discreteHamiltonian}

It is known that the preservation of first integrals (conservation
laws) in numerical work is of great importance (see, for
example,~\cite{Sam, Hairer}).   Therefore, it makes sense to
establish a discrete analog of the results presented for the
continuous Hamiltonian equations. {\bf An analogous discrete
framework would allow one to construct numerical schemes with first
integrals for various applied problems.}

\subsection{The discrete version of Hamiltonian action }

We will consider finite--difference equations and  discrete Hamiltonians
at some point $(t,{\bf q}, {\bf p})$  of a lattice.
Generally, the lattice in not regular.
The notations are clear from the following picture:

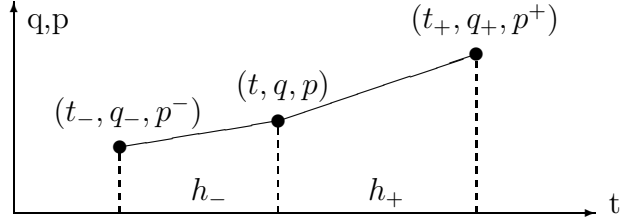
\begin{picture}(400,100)

\put(100,0){\begin{picture}(200,100)
\put(-10,10){\vector(1,0){220}}
\put(-10,10){\vector(0,1){80}} \put(30,35){\line(6,1){60}}
\put(90,45){\line(3,1){75}} \put(30,35){\circle*{5}}
\put(90,45){\circle*{5}} \put(165,70){\circle*{5}}

\put(-5,80){q,p}
\put(215,10){t}
\put(5,45){$(t_{-},q_{-},p^{-})$}
\put(74,55){$(t,q,p)$}
\put(140,80){$(t_{+},q_{+},p^{+})$}

\put(57,15){$h_{-}$} \put(124,15){$h_{+}$}
\multiput(30,10)(0,5){5}{\line(0,1){2.5}}
\multiput(90,10)(0,5){7}{\line(0,1){2.5}}
\multiput(165,10)(0,5){12}{\line(0,1){2.5}}
\end{picture}}

\end{picture}

To consider discrete equations we will need three points of
a lattice. Prolongation of Lie group operator~(\ref{symmetry})
for neighboring points $(t_-,{\bf q}_-, {\bf p}^- )$ and
$(t_+,{\bf q}_+, {\bf p}^+ )$ is the following:
\begin{equation}  \label{oper}
\begin{array}{c}
{\displaystyle
X =
\xi   { \partial \over \partial t } +
\eta ^i     { \partial \over \partial q ^i  } +
\zeta _i  { \partial \over \partial p_i }+
\xi _-   { \partial \over \partial t _- } +
\eta ^i  _-   { \partial \over \partial q ^i _-  } +
\zeta _i ^- { \partial \over \partial p_i ^-  }  } \\
\\
{ \displaystyle
+  \xi _+   { \partial \over \partial t _+ }
+ \eta ^i  _+   { \partial \over \partial q ^i _+  }
+ \zeta _i ^+ { \partial \over \partial p_i ^+  }
+ ( \xi _+  -  \xi  ) { \partial \over \partial {h _+} }
+ ( \xi   -  \xi  _- ) { \partial \over \partial {h _-} } , } \\
\end{array}
\end{equation}
where
$$
\xi _-  = \xi ( t _-, {\bf q}_-, {\bf p}^- ) ,
\qquad
{\eta_- ^i} =  {\eta ^i}  ( t _-, {\bf q}_-, {\bf p}^- ) ,
\qquad
{\zeta _i ^- }  = {\zeta ^i}  ( t _-, {\bf q}_-, {\bf p}^- ) ,
$$
$$
\xi _+  = \xi ( t _+, {\bf q} _+, {\bf p}^+ ) ,
\qquad
{\eta_+ ^i} =  {\eta ^i}  ( t _+, {\bf q} _+, {\bf p}^+ ) ,
\qquad
{\zeta _i ^+ }  = {\zeta ^i}  ( t _+, {\bf q} _+, {\bf p}^+ ) .
$$

Hamiltonian equations can be obtained by the variational principle
from the finite--difference functional
\begin{equation}  \label{actionHam}
\mathbb{H}_{h} = \sum^{}_{\Omega}  (  p _i ^+  ( q ^i _+ - q ^i  ) -
{\cal H} (t, t_+ , {\bf q}, {\bf p}^+ ) h_+  ) .
\end{equation}
Indeed, a variation of this functional along a curve $ q^i = \phi _i
(t)$, $p_i = \psi _i (t)$, $i =1, ..., n$ at some point $(t,{\bf
q},{\bf p})$ will effect only two terms of the
sum~(\ref{actionHam}):
\begin{equation} \label{j1}
\mathbb{H}_{h} = ... +   p _i   ( q ^i  - q ^i _-  )
-   {\cal H} (t _-, t  , {\bf q}_- , {\bf p} ) h_-
+   p _i ^+  ( q ^i _+ - q ^i  )
-   {\cal H} (t, t_+ , {\bf q}, {\bf p}^+ ) h _+ + ...
\end{equation}
Therefore, we get the following expression for the variation
\begin{equation} \label{j5}
\delta \mathbb{H}_{h}
=  \frac{\delta{\cal H}}{\delta p_i} \delta p_i
+  \frac{\delta{\cal H}}{\delta q^i} \delta q^i
+  \frac{\delta{\cal H}}{\delta t}  \delta t ,
\end{equation}
where $\delta q ^i  = {\phi_i ' } \delta t$,
$\delta p_i  = {\psi_i ' } \delta t$, $i = 1, ..., n$ and
\begin{equation} \label{Hamvariations}
\begin{array}{c}
{ \displaystyle \frac {\delta {\cal H}}{\delta p_i}
 =  q ^i -  q ^i _-
 - h _- { \partial {{\cal H}}  \over  \partial p_ i  } ^-   ,
\qquad
 \frac {\delta {\cal H}}{\delta q ^i}
 =  -  \left(   p_i ^+  -  {p_i}
 + h _+ { \partial {\cal H} \over  \partial q ^i }  \right) ,
\qquad i = 1, ..., n,  }\\
\\
{ \displaystyle  \frac{\delta{\cal H}}{\delta t}
= - \left(
 h _+   {\frac{\partial {\cal H} }{\partial t}} -  {\cal H}
+ h _-  {\frac{\partial {\cal H} } {\partial t }}^-  + {\cal H}^-
\right)
 } ,  \\
\end{array}
\end{equation}
where $ {\cal H} = {\cal H}(t,t _{+},{\bf q},{\bf p}^{+})$ and
${\cal H}^- =  {\cal H}(t _- ,t,{\bf q} _- ,{\bf p}) $.

For the stationary value of the finite--difference functional~(\ref{actionHam})
we obtain the system of $2n+1$ equations
$$
\frac {\delta {\cal H}}{\delta p_i} = 0 ,
\quad
\frac {\delta {\cal H}}{\delta q ^i} = 0 ,
\quad
i = 1, ..., n,
\qquad
\frac {\delta {\cal H}}{\delta t } = 0 .
$$
Thus, we arrive at the system of $2n+1$ equations
\begin{equation}   \label{Hamequations}
\begin{array}{c}
{\displaystyle { \dhp D } ( q ^i )
= {\frac{\partial { \cal H} }{\partial p _i ^+ }}  ,
\qquad
{ \dhp D } ( p_i )
= -{\frac{\partial {\cal H} }{\partial q ^i  }}  ,
\quad i = 1, ..., n, } \\
\\
{\displaystyle
 h _+   {\frac{\partial {\cal H} }{\partial t}} - {\cal H}
+ h _-  {\frac{\partial {\cal H}^- } {\partial t }} + {\cal H}^-    = 0} , \\
\end{array}
\end{equation}
which we will call {\it discrete Hamiltonian equations}. For
convenience we use the following
total shift (left and right) operators
and corresponding discrete differentiation operators:
$$
 { \dpm S} f(t) = f ( t _{\pm} ) ,
 \qquad
 { \displaystyle
 { \dpm D } = { {  \dpm S } -1 \over \pm h _{ \pm} } } .
$$
Let us note that the first $2n$ equations ~(\ref{Hamequations}) are
first-order discrete equations, which correspond to the canonical
Hamiltonian equations~(\ref{canonical})  in the continuous limit.
The last equation is of a second-order.
Its continuous counterpart
(see Remark 2.4) is automatically satisfied on the solutions of
canonical Hamiltonian equations.
In discrete case it defines the lattice on
which the canonical Hamiltonian equations are discretized. Being
second-order difference equation it needs one more initial  value
(first step of lattice) to state initial-value problem.
\par
It is interesting to note that the equations~(\ref{Hamequations})
can be obtained from discrete variational equations in Lagrangian
framework~\cite{Dorod1, Dorod2, Dorod3, dkw} with
the help of discrete Legendre transformation~\cite{Lall}.

\begin{remark}
Equivalent formulation can be considered for the
finite--difference functional
$$
\mathbb{H}_{h} = \sum^{}_{\Omega}  \left(  p _i  ( q ^i _+ - q ^i  )
- {\cal H} (t, t _+ , {\bf q} _+ , {\bf p} ) h _+  \right) .
$$
and a discrete Hamiltonian function $ {\cal H} (t, t _+ , {\bf q} _+ , {\bf p} ) $.
\end{remark}

\subsection{Invariance of the Hamiltonian action}

Let us consider the functional~(\ref{actionHam}) on some lattice,
given by equation
\begin{equation}  \label{mesh}
  {\Omega}(t,h _+, h _- ,  {\bf q} , {\bf p} )=0.
\end{equation}

\begin{definition}
We call a discrete Hamiltonian
function ${\cal H}$  considered on the mesh~(\ref{mesh}) {\it invariant} with
respect to a symmetry group generated by the operator~(\ref{oper}),
if the action~(\ref{actionHam}) considered on the mesh~(\ref{mesh})
is an invariant manifold of a group.
\end{definition}

\begin{theorem}   \label{discreteHinvaraince}
A Hamiltonian function considered together with the mesh~(\ref{mesh})
is invariant with respect to a group generated by
the operator~(\ref{oper}) if and only if the following conditions hold
\begin{equation}  \label{discreteinvar}
\begin{array}{c}
{ \displaystyle
\left.
 {\zeta _i ^+ }  \dhp D({q ^i})   +  {p_i^+} \dhp D( {  \eta}^i  )
 - X ({\cal H}) - {\cal H}  \dhp D( \xi )
 \right|
 _{ {\Omega}  =0 }
   = 0, }   \\
 \\
{ \displaystyle
\left.
 X{\Omega}(t,h_+, h _- , {\bf q} , {\bf p} )
\right|
 _{ {\Omega}  =0 }
=0 . } \\
\end{array}
\end{equation}
\end{theorem}

\noindent {\it Proof.}$\ $ The invariance condition follows directly
from  the action of $X$ on the functional:
$$
X \left(  \sum_{\Omega}  {p_i ^+ } (q ^i _+ - q ^i)
-  {\cal H}  h _+  \right)
= \sum_{\Omega} \left({\zeta _i ^+  }  \dhp D( q ^i )
+  {p_i ^+ } \dhp D( {\eta}^i ) -  X ({\cal H} )
 - {\cal H}  \dhp D( \xi ) \right) h _+ = 0 .
$$
It should be provided with the  invariance of a mesh, which is
obtained by the action of symmetry operator
on the mesh equation~(\ref{mesh}).
\hfill $\Box$

\subsection{Discrete Hamiltonian identity and discrete Noether--type theorem}

As in the continuous case, the invariance of a discrete Hamiltonian
on a specified mesh yields first integrals of discrete Hamiltonian equations.

\begin{lemma}   \label{lemmadiscreteidentity}
The following identity is true for any
smooth function
${\cal H} = {\cal H} (t, t _+ , {\bf q}, {\bf p} ^+ )$:
\begin{equation}  \label{Differ}
\begin{array}{c}
{ \displaystyle
{\zeta _i ^+  }   \dhp D( {q} ^i)  +  {p_i ^+  } \dhp D( { \eta}^i ) -
X ({\cal H} ) - {\cal H}   \dhp D( \xi ) \equiv \xi
\left( \dhp D({\cal H}^- )
 - {\partial {\cal H}  \over \partial t }
 - \frac{h_-}{h_+}{\partial {\cal H} \over\partial t }  ^- \right)  } \\
 \\
{ \displaystyle
- \eta^i    \left(\dhp D({p_i})
 +  { \partial {\cal H}  \over  \partial q ^i } \right)
 + {\zeta _i ^+ }\left(\dhp D({q} ^i)
 - {\partial {\cal H} \over \partial {p_i ^+} } \right)
+ \dhp D \left[   \eta^i {p_i} - \xi \left( {\cal H} ^- +
{h _-}{\partial {\cal H}  \over\partial t } ^- \right) \right] } \\
\end{array}
\end{equation}
\end{lemma}

\noindent {\it Proof.}$\ $
The identity can be established by direct
calculation.
\hfill $\Box$

\medskip

We call this identity the {\it discrete Hamiltonian identity}.
It allows us to state the following result.

\begin{theorem}   \label{dicretefirstintegral}
 The invariant with respect to symmetry operator~(\ref{oper})
discrete Hamiltonian equations~(\ref{Hamequations})
possess a  first integral
\begin{equation}  \label{discreteintegral}
{\cal I} =   \eta^i {p_i}
- \xi \left({\cal H}^- +  {h_-}{\partial {\cal H}  \over\partial t }  ^- \right)
\end{equation}
if and only if the Hamiltonian function is invariant with respect to
the same symmetry on the solutions of
equations~(\ref{Hamequations}).
\end{theorem}

\noindent {\it Proof.}$\ $
This result is a consequence of the identity~(\ref{Differ}).
 The invariance of the discrete Hamiltonian equations is needed
to guarantee the invariance of the mesh, which is defined by these
equations. \hfill $\Box$

\begin{remark}
Theorem~\ref{dicretefirstintegral} can be generalized for the
case of the divergence invariance of the Hamiltonian action, i.e.
\begin{equation}  \label{discretedivinvaraince2}
    \zeta _i  ^+   \dhp D( q ^i )   + p _i ^+  \dhp D( \eta ^i )
-  X ( {\cal H} ) - {\cal H}   \dhp D ( \xi )  = \dhp D( V ),
\end{equation}
where $V  = V ( t, {\bf q} ,{\bf p}) $. If this condition holds on
the solutions of the discrete Hamiltonian equations~(\ref{Hamequations}),
then there is a first integral
\begin{equation}  \label{discreteinteg1}
{\cal I } =  \eta^i {p_i}
- \xi \left({\cal H}^- +  {h _-} {\partial {\cal H} \over\partial t } ^- \right)
- V .
\end{equation}
\end{remark}

\begin{remark}
For discrete Hamiltonian equations with
Hamiltonian functions invariant with respect to time translations,
i.e. $ {\cal H} = {\cal H} (h _{+}, {\bf q}, {\bf p} ^{+})$, where
$h _+ = t_+ - t $, there is a conservation of energy
$$
{\cal E} = {\cal H}^-   +  {h _-} {\partial {\cal H}^-  \over\partial h_-  }
= {\cal H}   +  {h _+} {\partial {\cal H}  \over \partial h_+  }    .
$$
In this case the discrete Hamiltonian equations~(\ref{Hamequations})
are related to symplectic--momentum--energy
preserving variational integrations introduced
for discrete Lagrangian framework in~\cite{Kane}.
Note that ${\cal H}$ is not the discrete energy,
it has a meaning of a generating function for discrete
Hamiltonian flow.
\end{remark}

\subsection{Applications}

\subsubsection{Discrete harmonic oscillator}

The harmonic oscillator model is very important in physics.
A mass at equilibrium under the influence of any conservative force
behaves as a simple harmonic oscillator
(in the limit of small motions).
Harmonic oscillators are exploited in many manmade devices,
such as clocks and radio circuits.

Let us consider the one-dimensional harmonic oscillator
\begin{equation}   \label{original_oscilator}
\dot{q} = p ,  \qquad \dot{p} = - q  .
\end{equation}
This system of Hamiltonian equations is generated by the Hamiltonian
function
$$
H (t,q,p) = { 1 \over 2} ( q^2 + p ^2 ) .
$$

As a discretization of the equations~(\ref{original_oscilator})
we consider the application of the midpoint rule
\begin{equation}  \label{midpoint}
{ { q _+ - q  \over h_+ } =  { p + p _+   \over 2 }  } ,
\qquad
{ { p _+ - p  \over h_+ } = - { q + q _+   \over 2 } }
\end{equation}
on  a uniform mesh $ h _+ = h _- = h $.
The presented discretization can be rewritten as the following system of
equations
\begin{equation}  \label{Hamiltonian1a}
\begin{array}{c}
{\displaystyle { \dhp D } (q)
= {\frac{ 4 }{4 -  h _+ ^2 }}   \left( p _+  +  { h _+ \over 2 }  q \right) ,
\qquad
{ \dhp D } (p)
= - {\frac{ 4 }{4 -  h _+ ^2 }} \left( q + { h _+ \over 2 } p _+ \right) , } \\
\\
{\displaystyle  h_+ = h_-  .  } \\
\end{array}
\end{equation}
It can be shown that this system is generated by the discrete
Hamiltonian function
\begin{equation*}
{\cal H}  ( t, t _+ , q, p_+  )
= {\frac{ 2 }{4 - h_+ ^2 }} ( q^2 +  p_+   ^2 + h_+ q p_+ ) .
\end{equation*}
Indeed, the first and second equations of~(\ref{Hamequations}) are exactly
the same as those of~(\ref{Hamiltonian1a}).
The last equation of~(\ref{Hamequations}) takes the form
$$
- {\frac{ 2  (  4 +  h _+ ^2 ) }{ ( 4 -  h _+ ^2  ) ^2 }}  (  q^ 2 +  p _+ ^2 )
- {\frac{ 16  h _+  }{ ( 4 - h _+ ^2  ) ^2 }}  q p _+
+  {\frac{ 2  (  4 +  h _- ^2 ) }{ ( 4 - h _- ^2  ) ^2 }}  (  q _- ^ 2 + p ^2 )
+  {\frac{ 16  h _-  }{ ( 4 - h _-  ^2  ) ^2 }}  q _-  p
= 0 .
$$
Using the first and second equations,
we can rewrite it as
$$
\left(  - {\frac{ 2 }{  4 +  h _+  ^2 }  }
 +  {\frac{ 2 }{  4 +  h _-  ^2 }  } \right)
 ( q^2 + p^2 ) = 0 .
$$
Therefore, for the case $q^2 + p^2 \neq 0 $
this equation can be taken in
an equivalent form
$$
 h _+ = h _-  = h .
$$

The system of difference equations~(\ref{Hamiltonian1a})
admits,  in particular, the following symmetries
\begin{equation*}
X_1 = \sin(\omega t) {\frac{ \partial }{\partial q}}
+ \cos(\omega t) {\frac{\partial }{\partial p}} ,
\qquad
X_2 = \cos(\omega t) {\frac{ \partial }{\partial q}}
- \sin(\omega t) {\frac{ \partial }{\partial p}} ,
\end{equation*}
\begin{equation*}
X_3 = {\frac{ \partial }{\partial t}} ,
\qquad
X_4 = q {\frac{ \partial }{\partial q}} + p {\frac{ \partial }{\partial p}} ,
\qquad
X_5 = p {\frac{ \partial }{\partial q}} - q {\frac{ \partial }{\partial p}} ,
\end{equation*}
where
\begin{equation*}
\omega = {\frac{ \arctan ( h /2 ) }{h /2 }}.
\end{equation*}

For symmetry operators $X_1 $ and  $X_2$ we have the divergence
invariance conditions
$$
  \zeta   _+  \dhp D( q )    + p  _+  \dhp D( \eta  )
-  X ( {\cal H} ) - {\cal H}  \dhp D ( \xi  ) =  \dhp D (V)
$$
fulfilled on the solutions of equations~(\ref{Hamiltonian1a}) with
functions $V_1 = q \cos ( \omega t )$ and $V_2 = - q \sin ( \omega t
)$ respectively. Therefore, we obtain two corresponding first
integrals
\begin{equation}  \label{first_int1}
{\cal I} _1 = p \sin ( \omega t ) - q \cos ( \omega t ) ,
\qquad
{\cal I} _2 = p \cos ( \omega t ) + q \sin ( \omega t ) .
\end{equation}
Symmetry operator $X_3 $ satisfies the invariance condition
$$
  \zeta   _+  \dhp D( q )    + p  _+  \dhp D( \eta  )
-  X ( {\cal H} ) - {\cal H}  \dhp D ( \xi  ) =  0 .
$$
Thus,  we get the first integral
\begin{equation}  \label{first_int3}
{\cal I} _3 = - {\frac{ 4 }{4 - h _- ^2 }}
\left( {\frac{ 4 +  h _- ^2 }{4 -  h _- ^2 }} { \frac{  q _- ^ 2 + p ^2 }{2}}
+ {\frac{ 4 h _-  }{4 - h _- ^2 }} { q _-   p } \right)  .
\end{equation}
Using the first and second equations of~(\ref{Hamiltonian1a}),
we can simplify it as
\begin{equation*}
{\cal I} _3 = - {\frac{ 4 }{4 +  h _- ^2 }} {\frac{ q ^2 + p ^2 }{2}}
\end{equation*}
Since from the first integrals ${\cal I} _1$ and ${\cal I}_2$ we have conservation
\begin{equation*}
{\cal I} _1 ^2 + {\cal I} _2 ^2 = { q ^2 + p^2 } = \mbox{const} ,
\end{equation*}
it follows that we can take the third first integrals equivalently as
$$
\tilde{\cal I} _3 = h_-  .
$$

The three first integrals ${\cal I} _1$, ${\cal I} _2$, $\tilde{\cal I}_3$
are sufficient for integration of the system~(\ref{midpoint}).
We obtain the solution
$$
q = {\cal I} _2 \sin ( \omega t ) - {\cal I} _1 \cos ( \omega t ) ,
\qquad
p = {\cal I} _1 \sin ( \omega t ) + {\cal I} _2 \cos ( \omega t )
$$
on the lattice
$$
t_i = t_0 + i h, \qquad i = 0, \pm 1, \pm 2, ... , \qquad h =
\tilde{\cal I}_3 .
$$

\subsubsection{Modified discrete harmonic oscillator (exact scheme)}

The discrete harmonic oscillator of the preceding example follows
the same trajectory as the continuous harmonic oscillator, but with
a different velocity. This numerical error can be corrected by time
reparametrization. In this case we will get the exact discretization
of the harmonic oscillator, i.e. a discretization which gives the
exact solution of the underlying ODEs.

In this case the harmonic oscillator~(\ref{original_oscilator})
is discretized as
\begin{equation}  \label{Modmidpoint}
\begin{array}{c}
{ \displaystyle
{ { q _+ - q  \over h _+ } =  \Omega { p + p _+   \over 2 } } ,
\qquad
{ { p _+ - p  \over h _+ } = - \Omega { q + q _+   \over 2} } , } \\
\\
{ h _+ = h_- = h   ,} \\
\end{array}
\end{equation}
where
\begin{equation*}
\Omega = {\frac{\tan(h/2) }{h/2 }} .
\end{equation*}
represents a time reparametrization.
Similarly to the preceding example it can be shown that
this discrete model of the harmonic oscillator is
generated by the discrete Hamiltonian
\begin{equation*}
{\cal H}  (t, t _+,  q, p _+ )
= {\frac{ 2 \Omega }{4 - \Omega ^2 h _+ ^2 }}
 ( q^2 + p _+ ^2 + \Omega ^2 h _+  q p _+ ) .
\end{equation*}

The system of difference equations~(\ref{Modmidpoint}) admits the
following symmetries
\begin{equation*}
X_1 = \sin t  {\frac{ \partial }{\partial q}} + \cos t
{\frac{\partial }{\partial p}} , \qquad X_2 = \cos t  {\frac{
\partial }{\partial q}} - \sin t {\frac{ \partial }{\partial p}} ,
\end{equation*}
\begin{equation*}
X_3 = {\frac{ \partial }{\partial t}} ,
\qquad
X_4 = q {\frac{ \partial }{\partial q}} + p {\frac{ \partial }{\partial p}} ,
\qquad
X_5 = p {\frac{ \partial }{\partial q}} -q  {\frac{ \partial }{\partial p}} .
\end{equation*}
For symmetries $X_1 $ and $X_2 $,  which satisfy the divergence
invariance condition~(\ref{discretedivinvaraince2}) with functions $V_1 = q
\cos  t $ and $V_2 = - q \sin  t $, we obtain two first integrals
\begin{equation}  \label{first_int1_mod}
{\cal I} _1 = p \sin t  - q \cos t  ,
\qquad
{\cal I} _2 = p \cos t  + q \sin t  .
\end{equation}
The operator $X_3 $ satisfies the invariance condition~(\ref{discreteinvar})
and provides us with the first integral ${\cal I}_3$,
which (similarly to the preceding example)
can be taken in an equivalent form
\begin{equation}  \label{first_int3_mod}
\tilde{\cal I}_3 = h _- .
\end{equation}

The scheme~(\ref{Modmidpoint}) gives the exact solution of the
harmonic oscillator, which can be found with the help of first
integrals ${\cal I} _1$ and ${\cal I} _2$ as
$$
q = {\cal I} _2 \sin  t  - {\cal I} _1 \cos t  ,
\qquad
p = {\cal I} _1 \sin  t  + {\cal I} _2 \cos t .
$$
This discrete solution is given on the lattice
$$
t_i = t_0 + i h, \qquad i = 0, \pm 1, \pm 2, ... , \qquad
h = \tilde{\cal I}_3 .
$$

The exact schemes for two- and four-dimensional harmonic oscillators were used
in~\cite{Koz} to construct exact schemes for two- and three-dimensional Kepler motion
respectively.

\subsubsection{A non-linear motion}

We consider a difference analog of equations~(\ref{ca}), and choose

\begin{equation} \label{funty}
{\cal H} (t, t_+, q, p_+) =   { 1 \over 2}   \left(   p_+ ^2 + { 1 \over  q^2 } \right)  .
\end{equation}
Then, in accordance with (\ref{Hamequations}) we obtain the discrete
Hamiltonian equations:
\begin{equation}  \label{global2}
\begin{array}{c}
{ \displaystyle
\dhp D({q})   =  p _+ , \qquad \dhp D({p})  =  \frac{1}{q^3}, } \\
\\
{ \displaystyle   p_+ ^2 + { 1 \over  q^2 }
 =   p ^2 + { 1 \over  q_- ^2 }  } \\
\end{array}
\end{equation}

It is easy to check invariance conditions for $ {\cal H} $ with respect to
symmetry operators  $X_1$ and $X_2$, given in (\ref{opera}).
Application of Theorem~\ref{dicretefirstintegral} for these symmetries
yields first integrals
\begin{equation} \label{intJ1}
{\cal I} _1 =   - { 1 \over 2}   \left(   p ^2 + { 1 \over  q _-^2 } \right)  ,
\qquad
{\cal I} _2 = qp -  t  \left(   p ^2 + { 1 \over  q _- ^2 } \right)  .
\end{equation}
Therefore, the solution of the discrete system (\ref{global2})
satisfies the relation
$$
{\cal I} _2  =  qp  + 2t {\cal I} _1
$$
in all points of the lattice.

\section{Conclusion}
\label{conclusion}

The goal of the present paper is to present a method to find
first integrals of canonical Hamiltonian equations and to
establish a way to preserve Hamiltonian structure in
finite--difference schemes. To achieve this we use invariance of the
Hamiltonian action functional and its  relation to first integrals
of canonical Hamiltonian equations. The conservation properties of
the canonical Hamiltonian equations are based on the newly written
identity (called the Hamiltonian identity). This identity can be
viewed as a "translation" of the well--known Noether identity into
the Hamiltonian framework. The identity makes it possible to
establish one-to-one correspondence between invariance of the
Hamiltonian and first integrals of the canonical Hamiltonian
equations (the strong version of Noether's theorem).

The variational consequences of the Hamiltonian identity make it
possible to establish  necessary and sufficient conditions for the
canonical Hamiltonian equations to be invariant. These conditions
make it clear why not each symmetry of the Hamiltonian equations
provides a first integral.

The Hamiltonian version of Noether's theorem, formulated in the
paper, gives  a constructive way to find first integrals of the
canonical Hamiltonian equations once their symmetries are known.
This simple method does not require integration as it was
illustrated by a number of examples. In particular, we considered
equations of Kepler motion in various dimensions. The presented
approach gives a possibility to consider canonical Hamiltonian
equations and find their first integrals without exploiting the
relationship to the Lagrangian formulation
(see, for example,~\cite{Str}).

The approach developed for the continuous case
was applied to discrete Hamiltonian
equations, which can be obtained by variational principle from
finite--difference functionals. Similarly to the continuous case we
related invariance of discrete Hamiltonian functions to first
integrals of the discrete Hamiltonian equations. In particular,
energy conserving numerical schemes can be obtained as discrete
Hamiltonian equations generated by Hamiltonian functions invariant
with respect to time translations.
\par
The results presented in the paper can be used to find first
integrals of  continuous and discrete canonical Hamiltonian
equations. They also provide guidelines how to construct
conservative finite--difference schemes in Hamiltonian framework
that is important in numerical implementation.

\bigskip
\noindent{\bf \large Acknowledgments}
\medskip

The V.D.'s research was sponsored in part by the Russian Fund for
Basic Research under the research project no.~09-01-00610a. The
research of R.K. was partly supported by  the Norwegian Research
Council under contract no.~176891/V30.


\end{document}